\documentclass[final,5p,twocolumn]{elsarticle}
\journal{International Journal of Electrical Power and Energy Systems, February 2025}

\usepackage{amsthm} 
\usepackage[intlimits,fleqn]{amsmath}    
\usepackage[T1]{fontenc}
\usepackage[utf8]{inputenc}
\biboptions{sort&compress}
\usepackage{multirow}

\usepackage{bm}
\usepackage{epstopdf}
\usepackage{amssymb}
\usepackage{url}
\usepackage{enumitem} 
\usepackage{multirow}
\usepackage{hhline}
\usepackage{booktabs}
\usepackage{mathtools}
\usepackage{makecell}
\usepackage{algorithm}
\usepackage{comment}

\makeatother
\DeclareMathAlphabet\mathbfcal{OMS}{cmsy}{b}{n}



\usepackage{stackengine}

\newcommand{\m}{\boldsymbol}
\allowdisplaybreaks[4]
\usepackage[colorlinks = true,
linkcolor = blue,
urlcolor  = blue,
citecolor = blue,
anchorcolor = blue]{hyperref}

\usepackage[framemethod=TikZ]{mdframed}
\mdfdefinestyle{MyFrame}{%
	linecolor=black,
	outerlinewidth=1.25pt,
	roundcorner=1.25pt,
	innerrightmargin=5pt,
	innerleftmargin=5pt,}

\usepackage[noabbrev]{cleveref}

\usepackage{mathtools}

\DeclarePairedDelimiter\abs{\lvert}{\rvert}%
\DeclarePairedDelimiter\norm{\lVert}{\rVert}%

\makeatletter
\let\oldabs\abs
\def\abs{\@ifstar{\oldabs}{\oldabs*}}
\let\oldnorm\norm
\def\norm{\@ifstar{\oldnorm}{\oldnorm*}}
\makeatother

\usepackage[english]{babel}
\usepackage[utf8]{inputenc}
\usepackage[super]{nth}

\usepackage{graphicx}
\usepackage{float}
\usepackage[caption = false]{subfig}

\usepackage{array}
\usepackage{threeparttable}

\usepackage[english]{babel}
\usepackage[utf8]{inputenc}
\usepackage[super]{nth}

\usepackage{algpseudocode,float}
\usepackage{adjustbox}
\usepackage{graphicx} 
\usepackage{multirow}
\usepackage{tabularx}
\usepackage{booktabs}
\usepackage{rotating, makecell}
\usepackage{lipsum} 
\usepackage{hyperref}

\usepackage{etoolbox}
\makeatletter
\patchcmd{\ps@pprintTitle}
  {Preprint submitted}
  {Accepted}
  {}{}
\makeatother

\usepackage{amsmath}
\usepackage{amssymb}

\usepackage{csquotes}

\usepackage{lipsum}
\usepackage{lettrine}
\usepackage{eqnarray,amsmath}

\begin{document}
	\begin{frontmatter}
		\title{{\textbf{Unmasking Stealthy Attacks on Nonlinear DAE Models of Power Grids}}}
		
		\fntext[label2]{This work is supported by National Science Foundation under Grants 2230087 and 2404946.}
		
\author[inst1]{Abdallah Alalem Albustami}

\affiliation[inst1]{organization={Vanderbilt University},
	addressline={Civil and Environmental Engineering Department}, 
	city={Nashville},
	postcode={37235}, 
	state={TN},
	country={US}}
\ead{abdallah.b.alalem.albustami@vanderbilt.edu}

\author[inst1]{Ahmad F. Taha}
\ead{ahmad.taha@vanderbilt.edu}

\author[inst2]{Elias Bou-Harb}
\affiliation[inst2]{organization={Louisiana State University},
	addressline={Division of Computer Science and Engineering}, 
	city={Baton Rouge},
	postcode={70803}, 
	state={LA},
	country={US}}
\ead{ebouharb@lsu.edu}
		
\begin{abstract}
		Smart grids are inherently susceptible to various types of malicious cyberattacks that have all been documented in the recent literature. Traditional cybersecurity research on power systems often utilizes simplified models that fail to capture the interactions between dynamic and steady-state behaviors, potentially underestimating the impact of cyber threats. This paper presents the first attempt to design and assess stealthy false data injection attacks (FDIAs) against nonlinear differential algebraic equation (NDAE) models of power networks. NDAE models, favored in industry for  their ability to accurately capture both dynamic and steady-state behaviors, provide a more accurate representation of power system behavior by coupling dynamic and algebraic states. We propose novel FDIA strategies that simultaneously evade both dynamic and static intrusion detection systems while respecting the algebraic power flow and operational constraints inherent in NDAE models. We demonstrate how the coupling between dynamic and algebraic states in NDAE models significantly restricts the attacker's ability to manipulate state estimates while maintaining stealthiness. This highlights the importance of using more comprehensive power system models in cybersecurity analysis and reveals potential vulnerabilities that may be overlooked in simplified representations. The proposed attack strategies are validated through simulations on the IEEE 39-bus system.
		\end{abstract}

	\end{frontmatter}

\section{Introduction} \label{sec: intro}
In recent years, the pervasive adoption of information technology has elevated cybersecurity to a critical issue for contemporary power systems. The close interconnection between the physical and cyber dimensions of these systems suggests that a compromise in cybersecurity could pose a significant risk to their physical integrity \cite{Sridhar2012}. A particularly insidious threat within this domain is Stealthy False Data Injection Attacks (SAs).\footnote{We refer to these attacks by SAs, in contrast to the more verbose acronym SFDIAs used in the literature. We note that not all SAs are false data injection attacks, but the SAs considered in this paper are presumed to fall within the class of FDIAs.}  In particular, SAs have the capability to manipulate state estimation processes---that are instrumental in obtaining system-wide situational awareness and inform subsequent control actions---while evading conventional detection mechanisms, thereby threatening the stability and reliability of smart grids \cite{Liang2017}. 

Extensive research has been dedicated to the exploration of SAs against modern power systems, spanning the construction of valid attack strategies, assessing the impact of attacks on system stability, and developing defense and mitigation strategies \cite{Liang2017}.  The design and mitigation of SAs can be broadly categorized into three categories \cite{Reda2023}: \textit{(i)} those leveraging power flow equations (either DC or AC) to devise attack vectors \cite{Jia2012} and hence targeting steady-state grid conditions; \textit{(ii)} those exploiting the system's architectural vulnerabilities, including both centralized and decentralized attack schemes \cite{Lin2012, Kim2011}; \textit{(iii)} those predicated on the methodology of construction, ranging from attacks informed by complete or partial system topology---leveraging knowledge of the physical and logical layout of the power grid---as in \cite{Kosut2011a, Rahman2012a} to data-driven strategies that require no a priori knowledge of the grid's dynamics or parameters \cite{Nawaz2021, Kim2015, Chen2019}. 

In addressing SAs, two critical routines underpin the resilience of these infrastructures: state estimation (SE) and intrusion detection (ID) \cite{Liang2017}. State estimation is the process for inferring the grid's current state from measurements, essential for decision-making and monitoring. Intrusion detection works in tandem, discerning genuine system changes from erroneous or malicious data, hence safeguarding the state estimation process against potential anomalies \cite{Reda2023}. Significant efforts have been dedicated to developing and detecting stealthy data integrity attacks to compromise SE. This is a centerfold of most attack detection algorithms as most of them rely on computing residuals between predicted system state and estimated one. These attacks can be bounded, with limitations on their impact \cite{Guo2018}, or unbounded thereby offering wider scope for disruption \cite{Zhang2020}. Moreover, they can be meticulously tailored to create a worst-case scenario by exploiting the vulnerabilities of specific intrusion detectors \cite{Quinonez2020, Urbina2016}.

\noindent \textbf{Key Research Gap.} Despite extensive research, a critical gap exists in analyzing SAs against power systems modeled with Nonlinear Differential Algebraic Equations (NDAE). NDAE models, which are widely favored in industry for their ability to accurately represent both dynamic and steady-state behaviors of power systems \cite{sauer2017power}, offer several advantages over simplified linearized or Nonlinear ODE (NODE) models:

\begin{itemize}[leftmargin=*]
	\item More accurate representation of power system dynamics, capturing interactions between generator dynamic states (e.g., rotor angles, speeds, and transient voltages) and network algebraic constraints that model steady-state power flow behavior. This coupling allows for a more realistic simulation of system response to disturbances and control actions \cite{Gros2016, Nugroho2022}.
	\item Capacity to model expanded Phasor Measurement Unit (PMU) deployment across various bus types (e.g., load, renewable), extending beyond the generator-bus limitations of ODE models, potentially improving SE accuracy and fault detection \cite{Kazma2023a, Nadeem2023}.
	\item Continuous validation of state estimates against algebraic power flow and power balance constraints, improving the detection of measurement errors, model inaccuracies, and potential cyberattacks by identifying inconsistencies between the system's dynamic behavior and instantaneous power flow solutions.
\end{itemize}
These advantages can significantly impact SA analysis:
\begin{itemize}[leftmargin=*]
	\item \textit{Formulation:} NDAE models present attackers with a multifaceted challenge. They must design attack vectors that simultaneously: (1) keep detection residuals below established thresholds to evade conventional intrusion detectors, (2) ensure adherence to nonlinear algebraic constraints to maintain physical plausibility of the compromised system state, and (3) ensure the perturbed system remains observable and the state estimation process converges to a solution.		
	\item \textit{Detection and Mitigation:} The interdependence between dynamic and algebraic states in NDAE models enables multi-modal anomaly detection. Discrepancies between the temporal evolution of dynamic states and the instantaneous algebraic constraints can reveal subtle attack signatures not apparent in decoupled or simplified models.
	\item \textit{Attack Analysis:} NDAE models allow for a unified evaluation of how attacks on PMU measurement affect both dynamic states and steady-state variables, providing insights into both transient stability issues and steady-state constraint violations that may be overlooked in decoupled analyses. 
\end{itemize}

The lack of research in this area necessitates an in-depth investigation to gauge the susceptibility of NDAE-modeled power systems to SAs and the development of NDAE-specific detection and mitigation strategies. What follows is a detailed delineation of the literature that closely relates to this key gap of SA design and impact assessment on NDAE power system models.

\noindent \textbf{Relevant Literature.} Previous research on SAs against power systems has primarily focused on simplified models that may not fully capture the complex nonlinear dynamics and interactions present in real-world power grids. For instance, in \cite{Ding2019} and \cite{Ding2019b}, the authors investigate SAs using linearized \textit{descriptor} (another more mathematical phrase used to describe DAEs) systems. They employ a basic Bad Data Detection (BDD) scheme for intrusion detection and use an observer-based approach for SE. They define a perfect attack as one that satisfies certain detection avoidance and state divergence conditions, but they do not consider how the resulting state estimates must satisfy the algebraic constraints inherent in DAE models. Furthermore, their use of linearized models may not adequately capture the nonlinear behavior of power systems, potentially overlooking vulnerabilities that arise from the interaction between dynamic and algebraic states.

More recent studies, such as \cite{Lu2023}, \cite{Lu2022}, and \cite{en2022}, have shifted focus to FDIAs in nonlinear settings. However, these works primarily address static power flow models, optimizing attack sparsity and detection or leveraging deep learning for forecasting-based anomaly detection. They do not consider the coupled dynamic and algebraic components present in real systems, leaving a critical gap in understanding vulnerabilities arising from these interactions.

Studies like \cite{Hug2012} and \cite{Liang2016} have explored the inherent vulnerabilities of AC state estimation to false data injection attacks, with the former leveraging physical system properties for defense and the latter assessing physical disruptions from a bi-level optimization perspective. A significant body of work has also focused on constructing SAs against the AC state estimation process, which captures the algebraic constraints governing the steady-state behavior of power systems \cite{Jin2024,Rahman2013,Liu2021a,Mohammadi2022}. While these studies contribute to our understanding of system vulnerabilities under stealthy attack scenarios that exploit the algebraic constraints of steady-state behavior, they overlook dynamic system components, such as generator dynamics, and their interaction with the algebraic power flow equations. As a result, the impact of SAs designed using these approaches may not be fully representative of the actual system response when both dynamic and algebraic components are simultaneously considered.

To the best of the authors' knowledge,  this paper presents the first attempt to design and assess SAs against NDAE models of power systems that simultaneously account for generator dynamic behaviors and algebraic constraints that model steady-state power flow. 

\noindent \textbf{Investigated Research Questions.} This paper addresses this gap by investigating the following research questions: 
\begin{itemize}[leftmargin=*]
	\item \textit{Q1.} How does incorporating NDAE models influence the formulation, detection, and impact of SAs? Are there notable limitations and consequences when simplified models are employed to analyze and detect SAs in power systems?
	\item \textit{Q2.} How does the effectiveness of different mainstream intrusion detectors differ in detecting SAs for NDAE-modeled power systems? What improvements are necessary to enhance their detection capabilities?
	\item \textit{Q3.} How do NDAE-specific state monitoring algorithms, when coupled with dynamic intrusion detection methods, perform against tailored SAs?
\end{itemize}

\noindent\textbf{Paper Objectives and Contributions. } \;  This study examines attack strategies on NDAE-modeled power systems. We categorize attacks into two main strategies. \textit{(i)} \textit{Constraint-Unaware Attacks}: This strategy focuses on attacks that manipulate PMU measurements to remain undetected by intrusion detection systems but do not consider the adherence to the system's algebraic constraints, potentially leading to implausible physical conditions in the simulated attacks. \textit{(ii)} \textit{Constraint-Aware Attacks:} Here, stealthy attacks are designed to adhere to both detection evasion and the algebraic and operational constraints, ensuring attacks are both physically plausible and stealthy. The contributions of this paper are threefold:
\begin{enumerate}[leftmargin=*]
	\item Development of NDAE-aware SA strategies that ensure the resulting state estimates, encompassing both generator dynamics and steady-state variables, simultaneously satisfy the algebraic constraints of DAE models while remaining stealthy to intrusion detection. This approach captures the inherent coupling between dynamic and algebraic states, presenting a more comprehensive attack model than those considering only steady-state behavior. Specifically, we introduce an Iterative Constraint-Aware Attack Algorithm (ICAA), an efficient and scalable method for refining attack vectors to satisfy physical constraints while evading detection. ICAA begins with initial attack vectors designed to bypass detection thresholds (referred to as constraint-unaware attacks, as they do not consider physical system constraints) and systematically adjusts them to ensure physical plausibility, avoiding the need for direct solution of the NDAE model or complex optimization procedures, making ICAA computationally tractable for large-scale power systems.
	\item Evaluation of attack impacts on NDAE-specific state estimation and intrusion detection, comparing CUSUM and Chi-squared detector performance against the proposed attacks. This analysis reveals how NDAE models inherently constrain attack vectors compared to simplified power system representations, affecting both dynamic and algebraic state estimates.
	\item Quantitative assessment of NDAE model resilience against cyberattacks through case studies on the IEEE 39-bus system. We investigate trade-offs between constraint tolerance, false alarm rates, and attack detectability, providing insights into the security implications of using NDAE models for power system analysis and control.
\end{enumerate}

\noindent\textbf{Paper Organization.} The rest of the paper is organized as follows. Section \ref{sec:prelim} includes the preliminaries. Section \ref{sec:attacks} presents the attack strategies. Case studies are presented in Section \ref{sec:simulations}, and the paper is concluded in Section \ref{sec:conclusion}.

\section{Preliminaries}\label{sec:prelim}
This section summarizes the NDAE model, state estimation methods, and basic intrusion detectors critical for understanding and implementing the proposed algorithms. 

\subsection{Power Network NDAE Model}\label{sec:model}
Power systems can be modeled using a set of Nonlinear Differential Algebraic Equations (NDAE) that capture both the dynamic behavior of generators and the algebraic power flow constraints of the network. The NDAE model takes the following general form:
\begin{subequations}
	\begin{align}
		\textit{Gen. Dynamics:}\;\;	&{\boldsymbol{\dot{x}}}_d = \boldsymbol{A}_d \boldsymbol{x}_d + \boldsymbol{f}_d(\boldsymbol{x}_d, \boldsymbol{x}_a) + \boldsymbol{B}_d \boldsymbol{u}\\
				\textit{Flow Constraints:}\;\;		&\mathbf{0} = \boldsymbol{A}_a \boldsymbol{x}_a +  \boldsymbol{f}_a(\boldsymbol{x}_d, \boldsymbol{x}_a) + \boldsymbol{B}_a \boldsymbol{w}.	
	\end{align}
\end{subequations}

Here, $\boldsymbol{x}_d$ represents the dynamic state variables associated with generators, while $\boldsymbol{x}_a$ represents the algebraic state variables related to power flow in the network. The functions $\boldsymbol{f}_d$ and $\boldsymbol{f}_a$ capture the nonlinearities present in the generator dynamics and power flow equations respectively. Incorporating PMU measurements, represented by a vector \(\boldsymbol{y} \in \mathbb{R}^p\) of voltage and current phasors into the NDAE model, results in:
\begin{subequations}
	\begin{align}
		\boldsymbol{E} \dot{\boldsymbol{x}} &= \boldsymbol{A} \boldsymbol{x} +  \boldsymbol{f}(\boldsymbol{x}) + \boldsymbol{B}_u \boldsymbol{u} + \boldsymbol{B}_w \boldsymbol{w} + \boldsymbol{w}_p,\\
		\boldsymbol{y} &= \boldsymbol{C} \boldsymbol{x} + \boldsymbol{w}_m. ~\label{Eq:PMU measurement}
	\end{align}
\end{subequations}
The output matrix \(\boldsymbol{C} \in \mathbb{R}^{p \times n}\) links state vectors to PMU data via their geographic placements, with \(\boldsymbol{w}_m  \in \mathbb{R}^{p}\) and \(\boldsymbol{w}_p  \in \mathbb{R}^{n}\) representing random measurement and process noise.

For a complete and detailed description of the NDAE model, including the specific algebraic equations for power flow and power balance, please refer to \ref{appendixA}.

\subsection{NDAE State Estimation for Attack Detection} \label{sec:SE}
In the context of NDAE-modeled power systems, SE becomes particularly challenging due to the presence of both dynamic and algebraic states. Some methods in the literature simplify this task by reducing the model to NODE form or neglecting the algebraic constraints, compromising system representation fidelity \cite{Gros2016}. Incorporating algebraic constraints into the state estimation process for NDAE models can be accomplished through two primary approaches: a decoupled two-step approach and a coupled joint estimation approach. The decoupled approach first estimates algebraic variables using static SE techniques (such as Least Absolute Value estimation), then derives dynamic state estimates using dynamic estimation methods (such as the Extended Kalman Filter) \cite{Gol2014, Rouhani2018}. For the purpose of this work, we utilize the joint SE approach, which simultaneously estimates both dynamic and algebraic states, as it has demonstrated superior accuracy in capturing the interactions between dynamic and algebraic states in NDAE models, particularly in the presence of process and measurement noise, as well as uncertainties from loads and renewable energy sources \cite{Nadeem2023}.
\\
\\
\noindent{\textbf{Joint NDAE Observer.}}\label{sec:joint_observer} In \cite{Nadeem2023}, the authors introduce an observer that jointly estimates the dynamic and algebraic states of a NDAE-modeled power network using a Lyapunov-based, control-theoretic approach.	The observer and error dynamics are given by:
\begin{subequations} \label{eq:SE_joint}
	\begin{align}
		&\boldsymbol{E} \dot{\boldsymbol{\hat{x}}} = \boldsymbol{A} \boldsymbol{\hat{x}} + \boldsymbol{f}(\boldsymbol{x}) + \boldsymbol{B}_u \boldsymbol{u} +\boldsymbol{B}_w \boldsymbol{w} + 
		\boldsymbol{L}(\boldsymbol{y} - \boldsymbol{\hat{y}}), \label{eq:7a}	 \\
		&\boldsymbol{Z}\boldsymbol{\dot{e}} = (\boldsymbol{A} - \boldsymbol{L}\boldsymbol{C})\boldsymbol{e} + \Delta \boldsymbol{f} + (\boldsymbol{B_w} -\boldsymbol{ L}\boldsymbol{D_w})\boldsymbol{w}.  \label{eq:7b}
	\end{align}
\end{subequations}
The observer gain matrix \( \m L \) is determined through solving a convex semidefinite program problem that ensures that the error dynamics are bounded, hence guaranteeing that the state estimates are close to the actual system states under the presence of uncertainty from loads, renewables, and process and measurement noise. 

\subsection{Intrusion Detection}\label{sec:ID}
To analyze the impact of SAs on NDAE-modeled power systems, we employ two well-established residual-based intrusion detection methods: the Cumulative Sum (CUSUM) detector \cite{Gustafsson2001} and the Chi-squared ($\chi^2$) detector \cite{Plackett1983}. These detectors are chosen for their proven effectiveness in power system security and their compatibility with the utilized NDAE model \cite{Huang2016,Huang2011,Gol2015}.

\subsubsection{CUSUM Detection}
The Cumulative Sum (CUSUM) detector is a dynamic detection method that monitors changes in the cumulative sum of residuals, \( \m r \in \mathbb{R}^p \), defined as the differences between the state estimates from the SE step \eqref{eq:SE_joint} and the noisy measurements obtained from PMUs \eqref{Eq:PMU measurement}:
\begin{equation}
	\m r(t) = \m y(t) - \m C\hat{\m x}(t). \label{eq:residual}
\end{equation}
The CUSUM detection process is defined as:
\begin{equation} \label{eq:CUSUM}
	c_1 = 0, \;\; c_k = 
	\begin{cases}
		\max\left(0, c_{k-1} + z_k - b\right), & \text{if } c_{k-1} \leq \tau, \\
		0 \text{ and } \hat{k} = k-1, & \text{if } c_{k-1} > \tau.
	\end{cases}
\end{equation}

The parameters to be set are the bias \( b \in \mathbb{R}_{>0} \) and the threshold \( \tau \in \mathbb{R}_{>0} \). The output is the alarm time(s) \( \hat{k} \). An alarm is triggered when the test sequence \( c_{k-1} \) exceeds a predefined threshold $\tau$, and $c_k$ is set to zero. 
The parameters \( b \) and \( \tau \) are critical design choices that control the sensitivity and detection threshold of the CUSUM detector. While \cite{Murguia2016b} outlines a theoretical approach for tuning \( b \) and \( \tau \), this work employs a practical methodology based on analyzing historical residual data to tune these parameters.

The choice of the distance measure \( z_k \) depends on the configuration of the CUSUM detectors in the system. When using a scalar distance measure that aggregates information from all residuals, such as \( \m z_k = \m r_k^\top \m \Sigma^{-1} \m r_k \), the CUSUM parameters (\( b \), \( \tau \), and \( c_k \)) are also scalars. This centralized approach monitors all residuals collectively, treating the entire set of measurements as a single unit.
In contrast, if the distance measure is a vector, such as using the residual vector \( \m r_k \) itself or its absolute values \( |\m r_k| \), then the CUSUM parameters become vectors (\( b, \tau, c_k \in \mathbb{R}^p \)). This vectorized approach monitors each measurement's residual separately, allowing for different sensitivity for each measurement. This effectively provides a distributed detection capability within a single CUSUM framework as it allows to to detect localized anomalies to individual measurements. 

The CUSUM detector offers several advantages over traditional static detection methods, such as BDD. Unlike static methods which typically rely on single-shot residual analysis to flag anomalies, CUSUM dynamically tracks cumulative changes in residuals over time, making it more sensitive to subtle, persistent deviations introduced by stealthy attacks. Additionally, it is computationally efficient and straightforward to implement, as it requires only residual computations and simple iterative updates. This makes it a practical choice for real-time applications.

\subsubsection{Chi-squared ($\chi^2$) Detector}
The Chi-squared detector is a static detection mechanism designed to identify sudden anomalies in the residuals. Considering the residual vector \eqref{eq:residual}, the chi-squared procedure is defined as:\begin{equation} \label{eq:chi}
	z_k = \m r_k^\top \m \Sigma^{-1} \m r_k
\end{equation} 
An alarm is triggered when \( z_k \) exceeds a predefined threshold \( \alpha \). This threshold is set using the inverse regularized lower incomplete gamma function \( P^{-1}(\cdot,\cdot) \). Specifically, \( \alpha \) is calculated as \( \alpha = 2P^{-1}\left(\frac{n_y}{2}, 1 - \frac{1}{m}\right) \), where \( n_y \) is the number of independent measurements, and \( m \) represents the desired mean time between false alarms \cite{Murguia2016b}. The $\chi^2$ detector (also referred to as BDD) is widely used in FDIA research due to its effectiveness in identifying significant deviations indicative of malicious injections.

Next, we present two attack strategies designed to bypass the aforementioned intrusion detection systems, each accounting for different levels of system constraints.

\section{Attack Strategies}\label{sec:attacks}
In this section, we propose two attack strategies to evaluate whether the strict physical and operational constraints inherent in NDAE models---due to the coupling of dynamic and algebraic variables---make it more difficult for adversaries to evade SE and ID systems. We define two types of attacks: \textit{(i)} \textit{Constraint-Unaware Attacks (SCUAs)}, which aim to bypass ID methods without adhering to algebraic constraints. We formulate these attacks to explore their feasibility on the SE process of NDAE-modeled power systems when constraints are neglected, providing insights into applying similar attacks on simplified models that disregard grid physics. \textit{(ii)} \textit{Constraint-Aware Attacks (SCAAs)}, which maintain stealthiness against both ID systems and the NDAE constraints, ensuring the physical plausibility of the attacked state estimates. We point out here that these two problems have \textit{not} been studied in the literature for NDAE-modeled power systems, although they do sound intuitive and meaningful as outlined in Section \ref{sec: intro}. The effectiveness of both attack strategies is validated in Section \ref{sec:simulations}. Before detailing each strategy, we clarify key definitions used in attack formulation:

\begin{itemize}[leftmargin=*]
	\item Attack Vector (\(\boldsymbol{a}(k) \in \mathbb{R}^p\)): A vector that represents the manipulations applied to measurements at time $k$.
	\item Measurement Selection Matrix ($\boldsymbol{\Gamma} \in \mathbb{R}^{p \times p}$): A diagonal matrix where the $i$-th diagonal element is 1 if the $i$-th measurement is targeted by the attack, and 0 otherwise. This matrix defines which measurements are targeted by $\boldsymbol{a}(k)$. 
	\item Attack start time (\(k^*\)): Scalar defining the time step of attack initiation. 
	\item Attacked Measurements ($\boldsymbol{y}^*(k) \in \mathbb{R}^p$): The measurements obtained by adding the attack vector, scaled by the measurement selection matrix, to the true measurements at time step $k$, i.e., $\boldsymbol{y}^*(k) = \boldsymbol{y}(k) + \boldsymbol{\Gamma}\boldsymbol{a}(k)$.
\end{itemize}

\subsection{Stealthy Constraint-Unaware Attacks (SCUAs)} \label{SCUAs}
Stealthy Constraint-Unaware Attacks (SCUAs) are attacks designed to exploit vulnerabilities in ID systems by maximizing the injected false data while maintaining detector outputs at their predefined thresholds, without considering the algebraic and operational constraints of NDAE models. Below, we outline worst-case stealthy attack strategies for both detectors reproduced in Section \ref{sec:ID}, building on established methods from the literature \cite{Quinonez2020, Murguia2016}. 

\subsubsection{\textbf{SCUA formulation for CUSUM}} 
The CUSUM statistic under a stealthy injection attack considering a distance measure \( \m z_k = \m r_k^\top\m\Sigma^{-1}\m r_k \) and a single CUSUM statistic for all measurements is represented as:\footnote{While the NDAE model operates in continuous time, we use the discrete-time index $k$ in our attack formulations to reflect the discrete nature of digital measurements and control actions in power systems. Here, $k$ represents the sampling instant or time step of the measurement systems.}
\begin{equation}
	c_k = \max\left(0, c_{k-1} + (\m \Sigma^{-0.5}(\m r_k + \m a_k))^2 - b\right)
\end{equation}
An attack sequence, $\m a_k$, that can neutralize the residuals and maintain $c_k$ at its threshold can be defined as follows \cite{Murguia2016b}:
\begin{equation} 
\small	\m a_k = 
	\begin{cases}
		\m \Sigma^{\frac{1}{2}} \m\Gamma\left( \sqrt{\frac{\tau + b - c_{k-1}}{n}}, \ldots, \sqrt{\frac{\tau + b - c_{k-1}}{n}}\right)^\top - \m r_k,  \\ \text{if } k = k^* \\
		\m \Sigma^{\frac{1}{2}} \m\Gamma\left( \sqrt{\frac{b}{n}}, \ldots, \sqrt{\frac{b}{n}}\right)^\top - \m r_k,  \quad \text{if } k > k^*.
	\end{cases} \label{eq:CUSUM_SFDIA}
\end{equation}

The matrix $\m\Gamma$ selects the \( n \) measurements targeted by the attacker, where \( n \) corresponds to the number of ones on the diagonal of $\m\Gamma$ (i.e., number of selected measurements). This enables the attacker to spread the impact across any \( n \) selected measurements based on their access to PMU data, while still keeping the overall CUSUM statistic at the desired threshold.

In the case of considering a distance measure \( \m z_k = |\m r_k| \), where \( \m c_k = \max\left(0, \m c_{k-1} + |\m r_k + \m a_k| - \m b\right) \), the attack vector \( \m a_k \) can be designed to selectively target specific measurements to keep the CUSUM statistic \( c_k \) at desired levels. 
The attack sequence, \( a_k \) can be defined as follows \cite{Quinonez2020}:
\begin{equation}
	a_{k,i} = 
	\begin{cases}
		\pm (\tau_i + b_i - c_{k-1,i}) - \m r_{k,i}, & \text{if } k = k^*, \\
		b_i - \m r_{k,i}, & \text{if } k > k^*,
	\end{cases} \label{eq:CUSUM_dis_SFDIA}
\end{equation}
where \( i = 1, 2, \ldots, p \) represents each individual measurement. Here, the attacker must know the specific parameters of the CUSUM statistic for each targeted measurement. By selecting which measurements to attack and adjusting the magnitude of the injection based on this information, the attacker can maintain the CUSUM statistic \( c_k(i) \) for each targeted measurement \( i \) at its respective threshold \( \tau(i) \), thus implementing an effective SA in a distributed detection environment.

\subsubsection{\textbf{SCUA formulation for $\chi^2$}} 
To design a SA that bypasses the $\chi^2$ detector, the attack vector \( \m a_k \) must be crafted to ensure that the test statistic \( z_k \) does not exceed \( \alpha \). This can be achieved by controlling the total residual energy introduced by \( \m a_k \) as follows:	
\begin{equation} \label{eq:chi_a}
	\m a_k = \m\Sigma^{\frac{1}{2}} \m\Gamma \left( \sqrt{\frac{\alpha}{n}}, \ldots, \sqrt{\frac{\alpha}{n}}\right)^\top - \m r_k
\end{equation} 
This distribution allows the attacker to maintain the residual energy within the threshold \( \alpha \), thereby avoiding detection by the $\chi^2$ detector.

The attack vectors in \eqref{eq:CUSUM_SFDIA}, \eqref{eq:CUSUM_dis_SFDIA}, and \eqref{eq:chi_a} are designed to decieve the ID by ensuring the output remains consistently at its threshold. In these attacks, we assume that ID is done prior to the SE process, which means that it is used to check the measurements for anomalies before they are passed to the joint SE process, as this aligns with the paper's objective of quantifying the effectiveness of the attacks by analyzing their impact on SE accuracy. However, as the common practice would be to employ ID after SE, we also analyze this case, where residuals are computed using the current state estimates: $\boldsymbol{r}_k = \boldsymbol{y}_k - \boldsymbol{C}\hat{\boldsymbol{x}}_k$.
	For this case, we need to account for how the attack vector propagates through the state estimates. To avoid solving the complete NDAE observer or computing LMIs, we can approximate this propagation using a simple Euler discretization:$\hat{\boldsymbol{x}}_{k+1} = \hat{\boldsymbol{x}}_k + \Delta t(\boldsymbol{A}\hat{\boldsymbol{x}}_k + \boldsymbol{f}(\hat{\boldsymbol{x}}_k,\boldsymbol{u}) + \boldsymbol{L}(\boldsymbol{y}_k - \boldsymbol{C}\hat{\boldsymbol{x}}_k))$.
	Under attack, $\boldsymbol{y}^*_k = \boldsymbol{y}_k + \boldsymbol{a}_k$, the perturbed state estimate can be approximated as:
	\begin{equation}
		\hat{\boldsymbol{x}}^*_k \approx \hat{\boldsymbol{x}}_k - \Delta t\boldsymbol{L}\boldsymbol{a}_k
	\end{equation}
	This leads to the following residual expression: $
		\boldsymbol{r}^*_k = \boldsymbol{y}_k - \boldsymbol{C}\hat{\boldsymbol{x}}^*_k = (\boldsymbol{I} + \Delta t\boldsymbol{C}\boldsymbol{L})\boldsymbol{a}_k + \boldsymbol{r}_k, $
	Defining $\boldsymbol{M} = \boldsymbol{I} + \Delta t\boldsymbol{C}\boldsymbol{L}$, the attack vector can be designed as:
	\begin{equation}
		\boldsymbol{a}_k = \boldsymbol{M}^{-1}(\boldsymbol{\zeta} - \boldsymbol{r}_k)
	\end{equation}
	where $\boldsymbol{\zeta}$ is chosen based on the detector type as in \eqref{eq:CUSUM_SFDIA}, \eqref{eq:CUSUM_dis_SFDIA}, or \eqref{eq:chi_a}. Note that for small sampling times $\Delta t$, matrix $\boldsymbol{M}$ approaches identity, making both cases (ID before and after SE) nearly equivalent. In either case, these attack vectors serve as baselines that will be refined in the following section to satisfy physical constraints. 

	These strategies imply that once a SCUA is initiated, the detector's output remains consistently at its threshold. This highlights the importance of carefully tuning the threshold. If set too high, it may allow SAs to progressively diverge the state estimates from actual system states without detection. On the other hand, a threshold that is too low could result in numerous false alarms. 

	To launch these attacks, an adversary must have access to the measurements vector, state estimates, and the parameters of the intrusion detector. For the Chi-squared detector, this means knowing the threshold, $\alpha$. For the CUSUM detector, the attacker would need to know the value of $\tau$, $b$ and the CUSUM statistic from the prior step, $c_{k-1}$. The latter is particularly challenging, as $c_{k-1}$ is not typically communicated externally, which provides an inherent security advantage for the CUSUM detector over static ID methods \cite{Murguia2016b}. 

Although these assumptions may be idealized, they allow us to construct worst-case attack scenarios that serve as benchmarks for evaluating the robustness of more realistic and advanced strategies, as discussed in the next section.

Next, we propose a Stealthy Constraint-Aware Attack (SCAA) strategy that addresses the limitations of SCUAs by \textit{(i)} evading detection by intrusion detection systems, \textit{(ii)} ensuring state estimation converges to valid solutions, and \textit{(iii)} maintaining physically plausible state estimates that satisfy the dynamic and algebraic constraints of the NDAE model. This approach quantifies the more limited attack space in SCAAs compared to SCUAs, which do not consider these constraints. We introduce two methods for implementing SCAAs: an optimization-based approach that treats the problem as a constrained optimization and an algorithmic method leveraging closed-form solutions from SCUAs.

\subsection{Stealthy Constraint-Aware Attacks (SCAAs)}
In contrast to SCUAs that bypass detection without considering model constraints, Stealthy Constraint-Aware Attacks (SCAAs) are designed to evade detection while adhering to the algebraic and operational constraints of NDAE-modeled power systems.
These constraints, denoted as $\mathcal{C}$, include the algebraic power flow \eqref{eq:PG}--\eqref{eq:Qg} and power balance equations \eqref{eq:realbal}--\eqref{eq:compbal}, as well as the operational limits on generator outputs \eqref{eq:con5}--\eqref{eq:con6}, bus voltage magnitudes \eqref{eq:con7}, and transmission line flows \eqref{eq:con8}--\eqref{eq:con9}, as follows:

\vspace{-0.3cm}
{
	\small
	\begin{subequations}\label{eq:constraints}
		\begin{align}			
			&P_{Gi} = \frac{1}{x'_{di}} E'_{qi}v_i \sin(\delta_i - \theta_i) - \frac{x_{qi} - x'_{di}}{2x'_{di}x_{qi}}v_i^2 \sin(2(\delta_i - \theta_i)) \label{eq:PG} \\
			&Q_{Gi} = \frac{1}{x'_{di}} E'_{qi}v_i \cos(\delta_i - \theta_i) - \frac{x'_{di} + x_{qi}}{2x'_{di}x_{qi}}v_i^2  \nonumber \\
			&\phantom{=} - \frac{x_{qi} - x'_{di}}{2x'_{di}x_{qi}}v_i^2 \cos(2(\delta_i - \theta_i)) \label{eq:Qg} \\
			&P_{Gi} + P_{Ri} - P_{Li} = \sum_{j=1}^{N} v_i v_j (G_{ij} \cos \theta_{ij} + B_{ij} \sin \theta_{ij}) \label{eq:realbal} \\
			&Q_{Gi} + Q_{Ri} - Q_{Li} = \sum_{j=1}^{N} v_i v_j (G_{ij} \cos \theta_{ij} - B_{ij} \sin \theta_{ij}) \label{eq:compbal},\\
			&P_{Gi}^{\min} \leq P_{Gi} \leq P_{Gi}^{\max} \label{eq:con5},\\
			&Q_{Gi}^{\min} \leq Q_{Gi} \leq Q_{Gi}^{\max} \label{eq:con6},\\
			&V_i^{\min} \leq V_i \leq V_i^{\max} \label{eq:con7},\\
			&S_{f_i} \leq F_{\max} \label{eq:con8},\\
			&S_{t_i} \leq F_{\max}\label{eq:con9},
		\end{align}
	\end{subequations}
}
where $S_{f_i}$ and $S_{t_i}$ represent the apparent power flows at the from and to ends of a transmission line, respectively, and $F_{\max}$ denotes the line's maximum rating. These variables are functions of active and reactive powers, voltage magnitudes at the corresponding buses, and the line admittance. For a detailed description of the variables and their specific formulations, refer to  \ref{appendixA}.

	SCAAs aim to manipulate PMU measurements to introduce disturbances that can  impact the state estimation process, all while ensuring that the attacked state estimates comply with both the algebraic and operational constraints in $\mathcal{C}$.\\
To ensure that the SCAA approach computationally tractable for larger systems, we start by defining an attack zone that captures how perturbations propagate through the network topology. Algorithm \ref{algo:zone} in \ref{appendixB}  determines this zone by analyzing the network's admittance matrix structure and bus classifications. Starting from the targeted measurement buses, the algorithm propagates through zero-injection buses while adding non-zero injection buses to the boundary of the attack zone. This approach captures all buses that could be affected by the attack while maintaining a tractable problem size.
	The algorithm also identifies the specific state variables that must be considered within this zone, including both dynamic states of generators and algebraic states of buses that either lie within or are directly connected to the attack zone. This targeted approach allows us to enforce constraints only where they are physically relevant, significantly reducing computational complexity compared to considering the entire system.

To formally define the SCAA strategy, we formulate the following optimization problem:
\begin{subequations} \label{eq:OP_SFDIA}
	\begin{align}
		\underset{\boldsymbol{a}_k}{\text{maximize}} &\quad \sum_{i=1}^{p} |\boldsymbol{a}_k| \label{eq:objective} \\ \text{subject to} & \quad \hat{\boldsymbol{x}}^*_{\mathcal{A}}(k) = \mathcal{F}(\hat{\m x}_{\mathcal{A}}(k-1), \m y^*(k)), \label{eq:SE_att} \\ 
		& \quad \left\lvert  \sum \boldsymbol{g}_L(\hat{\boldsymbol{x}}^*_{\mathcal{A}}(k)) \right\rvert \leq \zeta, \label{eq:equ_att} \\
		& \quad \boldsymbol{h}(\hat{\boldsymbol{x}}^*_{\mathcal{A}}(k)) \leq \boldsymbol{0}, \label{eq:ineq_att} \\ 
		& \quad \mathcal{D}(\hat{\m x}^*_{\mathcal{A}}(k), \m y^*(k)) \leq \gamma \label{eq:ID_att}, \\
		& \quad \m y^*(k) = \m{y}(k) + \m\Gamma \m{a}_k
	\end{align} 
\end{subequations}

The objective function \eqref{eq:objective} aims to maximize the sum of the absolute values of the attack vector components, representing the maximum perturbation an attacker can inject into the selected measurements to disrupt the SE process. The stealthiness and physical plausibility of the attack are ensured through a series of constraints imposed on the manipulated measurements and the resulting altered state estimates.

Constraint \eqref{eq:SE_att} represents computing the attacked state estimates resulting from perturbed measurements within the attack zone. This ensures that the optimization considers how the attack will affect the SE outcome.
Constraints \eqref{eq:equ_att} and \eqref{eq:ineq_att} ensure physical plausibility by enforcing power system algebraic and operational constraints {on the states in the set $\mathcal{S}$ defined by Algorithm \ref{algo:zone} in \ref{appendixB}. Lastly, constraint \eqref{eq:ID_att} ensures that the ID output $\mathcal{D}$ remains below the detector's threshold $\gamma$. This constraint can represent either \eqref{eq:CUSUM} for the CUSUM statistic, or \eqref{eq:chi} for the $\chi^2$ statistic, depending on the ID method in use.

To make the proposed SCAA optimization formulation computationally feasible, we linearize the algebraic constraints in \eqref{eq:equ_att} using a first-order Taylor expansion to enable more tractable optimization. We ensure that the linear approximation remains valid by periodically updating the linearization points during the attack. The linearized algebraic constraints are expressed as $$\boldsymbol{g}_L(\hat{\boldsymbol{x}}^*(k)) = \boldsymbol{g}(\hat{\boldsymbol{x}}_0) + \nabla\boldsymbol{g}(\hat{\boldsymbol{x}}_0)(\hat{\boldsymbol{x}}^*(k) - \hat{\boldsymbol{x}}_0).$$

Function \( \mathcal{F}(\cdot) \) in \eqref{eq:SE_att}, which is used to compute the attacked state estimates, can be defined in several ways. One option is to assume the attacker has access to a joint observer, allowing them to estimate the attacked states by solving a modified state estimation problem similar to \eqref{eq:SE_joint}. Alternatively, the attacker could update their state estimates dynamically using a pseudo-inverse approach, \( \hat{\boldsymbol{x}}_a^*(k) = \hat{\boldsymbol{x}}_a(k-1) + \boldsymbol{C}^\dagger\boldsymbol{y}^*(k)\), assuming minimal changes between time steps. This approach works well when the power system's state evolves slowly, and the pseudo-inverse \( \boldsymbol{C}^\dagger \) exists and is reliable. However, practical limitations, such as measurement redundancy or errors in system topology, may affect the rank of \( \boldsymbol{C} \), reducing the effectiveness of this method.

While the optimization-based approach provides a formal framework for SCAAs, we propose an alternative, computationally efficient method called the Iterative Constraint-Aware Attack Algorithm (ICAA). This algorithm leverages the closed-form solution of SCUAs and iteratively scales it down to satisfy both intrusion detection and physical constraints of the NDAE system.

\subsection{Iterative Constraint-Aware Attack Algorithm (ICAA)}
The key idea behind ICAA is to start with the maximum possible attack magnitude that satisfies the intrusion detection constraint. If this initial attack vector satisfies all system constraints, it is immediately applied. Otherwise, the algorithm iteratively reduces the attack magnitude until all constraints are met. This approach offers several advantages. It is computationally efficient, avoiding complex optimization solvers and making it suitable for real-time applications. The iterative nature ensures both intrusion detection and physical constraints are always satisfied, adapting to the current system state and potentially allowing larger attacks when constraints are less binding. Unlike the optimization approach, ICAA doesn't require constraint linearization, avoiding approximation errors and enhancing accuracy in highly nonlinear systems. This method balances attack impact with constraint satisfaction, providing an alternative to optimization-based techniques. The ICAA is presented in Algorithm \ref{alg:ICAA}.

\begin{algorithm}[http]
	\caption{Iterative Constraint-Aware Attack}\label{alg:ICAA}
	\begin{algorithmic}[1]
		\Require $\boldsymbol{y}(k)$, $\hat{\boldsymbol{x}}(k)$, $\alpha$ (for $\chi^2$), $\tau$, $b$ (for CUSUM), $\boldsymbol{\Gamma}$, $\zeta$, $\beta$, $N_{\text{max}}$, detector type
		\Ensure $\boldsymbol{a}_k$
		\State Initialize $\boldsymbol{a}_k = \boldsymbol{0}$, set $\gamma \gets \alpha$ ($\chi^2$) or $\tau$ (CUSUM)
		\State $\boldsymbol{a}_k(\boldsymbol{\Gamma}) \gets$ attack vector from \eqref{eq:CUSUM_SFDIA} or \eqref{eq:chi_a} based on ID type
		\State $\boldsymbol{y}^*(k) \gets \boldsymbol{y}(k) + \boldsymbol{a}_k$
		\State $\hat{\boldsymbol{x}}^*(k) \gets \mathcal{F}(\boldsymbol{y}^*(k), \hat{\boldsymbol{x}}(k-1))$
		\For{$i = 1$ to $N_{\text{max}}$}
		\If{$|\sum \boldsymbol{g}(\hat{\boldsymbol{x}}^*(k))| \leq \zeta$ \textbf{and} $\boldsymbol{h}(\hat{\boldsymbol{x}}^*(k)) \leq \boldsymbol{0}$ \textbf{and} $\mathcal{D}(\m y^*(k), \hat{\m x}^*(k)) \leq \gamma$ (\eqref{eq:CUSUM} or \eqref{eq:chi})}
		\State \textbf{break} \Comment{All constraints satisfied}
		\EndIf
		\State Scale $\boldsymbol{a}_k$ by $(1-\beta)$ \Comment{Reduce attack magnitude}
		\State $\boldsymbol{y}^*(k) \gets \boldsymbol{y}(k) + \boldsymbol{a}_k$
		\State $\hat{\boldsymbol{x}}^*(k) \gets \mathcal{F}(\boldsymbol{y}^*(k), \hat{\boldsymbol{x}}(k-1))$
		\EndFor
		\If{$i = N_{\text{max}}$}
		\State $\boldsymbol{a}_k \gets \boldsymbol{0}$ \Comment{Attack infeasible, revert to no attack}
		\EndIf
		\State \Return $\boldsymbol{a}_k$
	\end{algorithmic}
\end{algorithm}

The algorithm takes as input the current measurements $\boldsymbol{y}(k)$, detector parameters, attack selection matrix $\boldsymbol{\Gamma}$, constraint tolerance $\zeta$, scaling factor $\beta$, maximum iterations $N_{\text{max}}$, and detector type.
It starts by initializing the attack vector to zero and setting the detection threshold $\gamma$ based on the detector type. The initial attack vector is then computed using the SCUA formulation for the chosen detector. The algorithm then enters a loop where it iteratively checks if the current attack vector satisfies all constraints.
The function $\mathcal{F}(\cdot)$ in line 4 computes the attacked state estimates. This can be implemented as a single step of the joint observer \eqref{eq:SE_joint} or using a pseudo-inverse approach.
If all constraints are satisfied, the algorithm terminates and returns the current attack vector. Otherwise, it scales down the attack vector by a factor of $(1-\beta)$ and recomputes the attacked measurements and state estimates.
This process continues for a maximum of $N_{\text{max}}$ iterations. If no feasible attack is found within these iterations, the algorithm reverts to no attack.

The proposed algorithm is highly scalable, irrespective of the size and order of the NDAE model. A key advantage of the approach is that it does not require solving the NDAE model at each iteration, which is typically computationally expensive. Instead, it begins with attacks designed as upper bounds based on the intrusion detection (ID) and state estimation (SE) parameters (referred to as SCUAs) and iteratively refines them. Since the initial SCUA attack vectors are computed in constant time, $O(1)$, and the algorithm's refinement process involves straightforward scaling operations and constraint checks, its overall computational complexity remains $O(1)$ as well. This  ensures the algorithm's practical for real-time implementation even in systems with high-dimensional state spaces or complex dynamics.

Regardless of the SCAA formulation the attacker chooses, it requires them to have comprehensive knowledge of both the intrusion detection methods (e.g., \eqref{eq:CUSUM_SFDIA} or \eqref{eq:chi_a}) and the NDAE system. Specifically, the attacker must possess detailed information about the NDAE model's algebraic and dynamic equations, operational limits, and the structure of the SE process. Refer to Tab.~\ref{tab:knowledge_requirements} for the specific knowledge requirements for the proposed attack strategies. While these assumptions might be idealistic, they align with the literature on constructing SAs against AC SE, and we leverage them not merely to formulate a perfect attack, but rather to explore the efficacy of NDAE models in offering enhanced protection against tailored SAs. Specifically, we compare the impact of SCUAs and SCAAs on NDAE-specific SE process to assess whether the inherent coupling in these models reduces the attack space and increases resilience---an aspect that has been overlooked in studies using simplified models.

\begin{figure}[t]
	\centering
	\includegraphics[width=0.95\columnwidth]{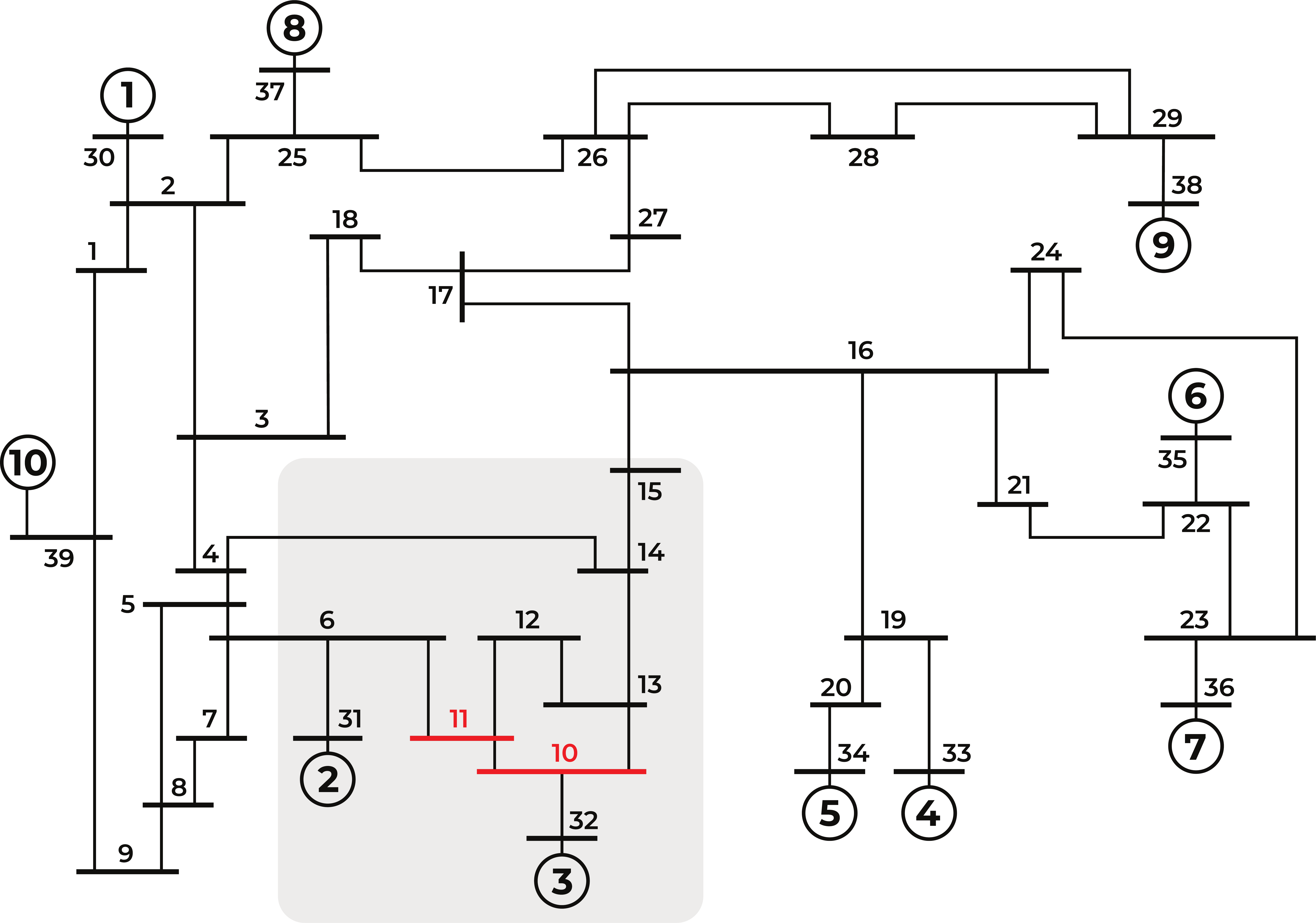}
	\caption{Single-line diagram of the IEEE 39 Bus Network.} 
	\label{fig:IEEE39bus}
\end{figure}

	\begin{table}[t] 
	\centering
	\caption{Knowledge Requirements for SCUAs and SCAAs/ICAAs}
	\label{tab:knowledge_requirements}
	\renewcommand{\arraystretch}{1.5} 
	\setlength{\tabcolsep}{10pt} 
	\resizebox{\linewidth}{!}{%
		\begin{tabular}{p{3.7cm}|p{2.5cm}|p{3cm}}
			\hline
			\textbf{Knowledge Area} & \textbf{SCUA} & \textbf{SCAA/ICAA} \\ \hline\hline
			\textbf{ID Parameters} & \multicolumn{2}{p{6cm}}{Detection thresholds (\(\alpha, \tau, b\)) and residual covariance matrix (\(\boldsymbol{\Sigma}\))} \\ \hline
			\textbf{SE Parameters} & \multicolumn{2}{p{6cm}}{Observer gain matrix (\(\boldsymbol{L}\)) and measurement matrix (\(\boldsymbol{C}\))} \\ \hline
			\textbf{System Topology} & Not required & Y-Bus matrix (\(\boldsymbol{Y}_{\text{bus}}\)) and knowledge of grid topology \\ \hline
			\textbf{Operational Constraints} & Not considered & Power flow equations and operational limits within the attack zone $\mathcal{A}$\\ \hline
		\end{tabular}%
	}
\end{table}

\section{Case Studies}\label{sec:simulations}	\subsection{Simulation Setup and Parameters}
We assess the feasibility and impact of the proposed SA strategies on the fourth-order NDAE model discussed in Section \ref{sec:model}. We conduct case studies using the IEEE 39-bus system. System parameters are obtained from the MATPOWER toolbox \cite{Zimmerman2011}. For the 39-bus system, following \cite{Rimal2022}, we install PMUs at buses 2, 6, 9, 10, 13, 14, 17, 19, 20, 22, 23, 25, and 29 to ensure complete observability. 

The attack scenarios are initiated at $t = 15$ sec, allowing the system to reach steady-state conditions prior to the attack. The simulations are run for a total of $30$ sec to capture the short-term and long-term impacts of the attacks on the power system dynamics, SE process, and ID performance. 

The threshold of the $\chi^2$ detector is set according to \( \alpha = 2P^{-1}\left(\frac{58}{2}, 1 - \frac{1}{1000}\right) \) = 99.175 For a CUSUM detector that uses a combined distance measure $\m z_k = \m r^\top \m \Sigma^{-1} \m r$, the threshold is set to $\tau = 116.28$ and the bias it set to $b = 50.14$ to establish the same false alarm rate of the $\chi^2$ detector. For a CUSUM that uses a vectorized distance measure, $\m z = |\m r|$, there are 58 thresholds and 58 bias parameters, omitted for brevity.
The parameters of the ICAA strategy (Algorithm \ref{alg:ICAA}) are set to $N_{max} = 100$, $\beta = 0.01$, and $\zeta = 0.22$. We discuss how the choice of $N_{max}$ and $\beta$ affects the attack impact, and how the choice of $\zeta$ for a system operator affects attack detection through constraint validation.

Renewables are integrated into the NDAE model by modifying the algebraic power flow equations to account for renewable disturbances. Specifically, the real and reactive power from renewables (\(P_R\) and \(Q_R\)) are included as components of the algebraic variables, \(q(t)\), which represent system-wide power flows. The algebraic constraints of the NDAE model are given by: $0 = A_a x_a + F_a f_a(x_d, x_a) + B_a q,$ where \(q = [P_R^\top, Q_R^\top, P_L^\top, Q_L^\top]^\top\), and \(P_R\) and \(Q_R\) represent the power contributions from renewables. To incorporate variability, we model renewable power injections as: $q(t) = \bar{q} + \Delta q(t),$ where \(\bar{q}\) represents the nominal steady-state power injection and \(\Delta q(t)\) captures time-varying disturbances, including stochastic variations. These disturbances are modeled as Gaussian noise with zero mean and variance proportional to the renewable generation capacity \cite{Nadeem2023}.
	
Fig. \ref{fig:IEEE39bus} illustrates the IEEE 39-bus system under study, highlighting the buses targeted in the attack scenarios (10 and 11), marked in red. The attack zone, calculated using Algorithm \ref{algo:zone}, is determined to be $\mathcal{A} = \{5, 6, 10, 11, 12, 13, 31, 32\}$, as highlighted in the gray box. While these buses define the primary attack zone for most simulations, alternative sets of target buses may be occasionally used, and specific changes are noted when applicable.

\subsection{Impact of SAs on State Estimation Accuracy}
We compare the impact of the two proposed SA strategies (SCUAs and SCAAs) on the SE process in Fig. \ref{fig:Impact on SE}. For intrusion detection, a CUSUM detector is employed with a combined distance metric for all residuals. The Root Mean Square Error (RMSE)  was chosen because of its sensitivity to both small and large deviations, making it particularly suitable for capturing the impact of stealthy attacks.

The RMSE value under normal operation is 0.3178. SCUAs result in a significant increase in this value to 1.2437. This can also be observed in the deviations in both algebraic and dynamic states. In contrast, SCAAs demonstrate more subtle impacts on state estimates, with an RMSE of 0.36286. The frequency, voltage magnitude, and rotor angle estimates under SCAAs remain closer to their true values, with only minor fluctuations. 

The NDAE model and NDAE-specific SE techniques allow us to observe how measurement manipulation immediately propagates to dynamic generator state estimates. This represents a significant advantage over studies focusing solely on AC SE processes, where only steady-state estimates can be monitored. The simultaneous simulation and estimation of dynamic and algebraic states in the NDAE framework provides attackers with less opportunity to manipulate measurements while ensuring that resulting state estimates satisfy algebraic constraints and evade intrusion detection.
This increased difficulty for attackers likely stems from the interaction terms between dynamic and algebraic variables in equations \eqref{eq:PG} and \eqref{eq:Qg}. These interactions significantly limit the feasible attack space, as changes in one state variable have cascading effects on others that must be accounted for to maintain plausibility. This demonstrates how NDAE models inherently provide greater cybersecurity resilience compared to decoupled or simplified power system representations.

In Fig.~\ref{fig:MAE}, we further illustrate the impact of SCUAs and SCAAs on the state estimation process using three metrics: the Mean Absolute Error (MAE) over time and the absolute error dynamics for selected states under both attack strategies. The first subplot highlights that SCAA/ICAA consistently achieves lower MAE values compared to SCUA, reflecting the reduced impact of attacks when algebraic and operational constraints are satisfied. The second and third subplots display the absolute error dynamics for six representative dynamic states of generator 1 under SCUA and SCAA/ICAA, respectively. These plots emphasize how constraint-aware attacks limit the attacker’s ability to manipulate the system, as evidenced by the smaller deviations in state estimates. This underscores the fact that requiring the attacker to satisfy the physical and algebraic constraints of the NDAE model significantly reduces their room for manipulation, ultimately enhancing the system's resilience against stealthy attacks.

\begin{figure}[http]
	\centering
	\includegraphics[width=\columnwidth]{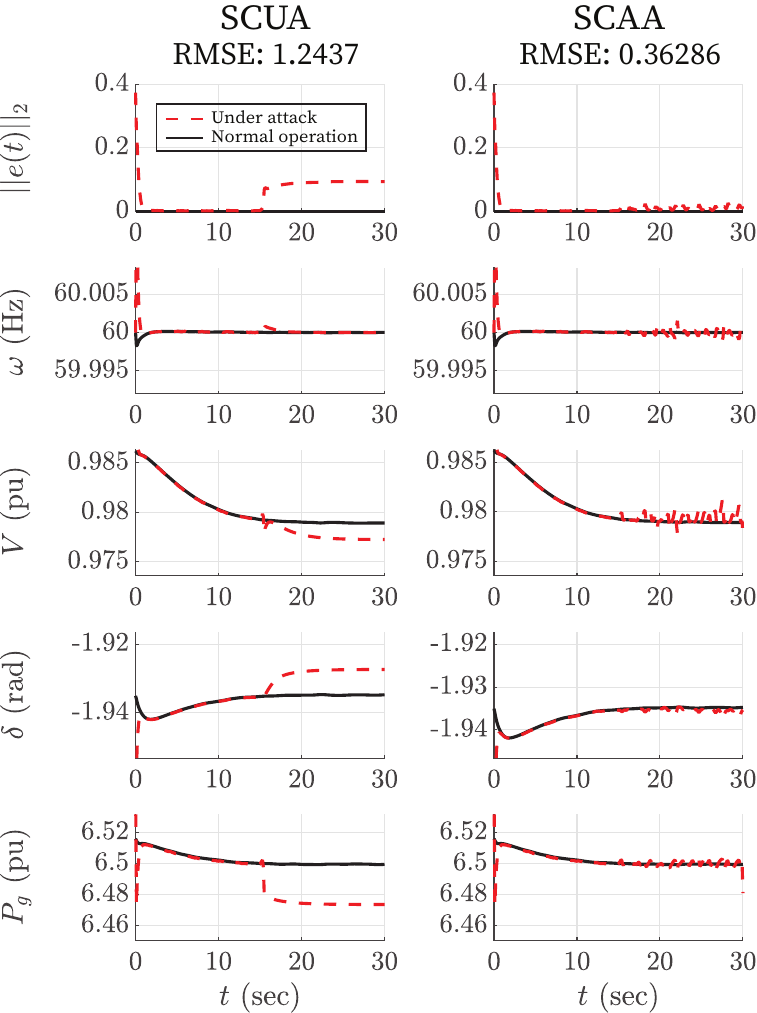}
	\caption{Comparison of power system states of generator 2 under SCUAs (left column) and SCAAs (right column). From top to bottom: (1) SE error norm, (2) Frequency, (3) Bus voltage magnitudes, (4) Generator rotor angle and (5) Generator active power outputs. Solid black lines represent true states, while red dashed lines show estimated states under attack.} 
	\label{fig:Impact on SE}
\end{figure}
\begin{figure}[http]
	\centering
	\includegraphics[width=0.9\columnwidth]{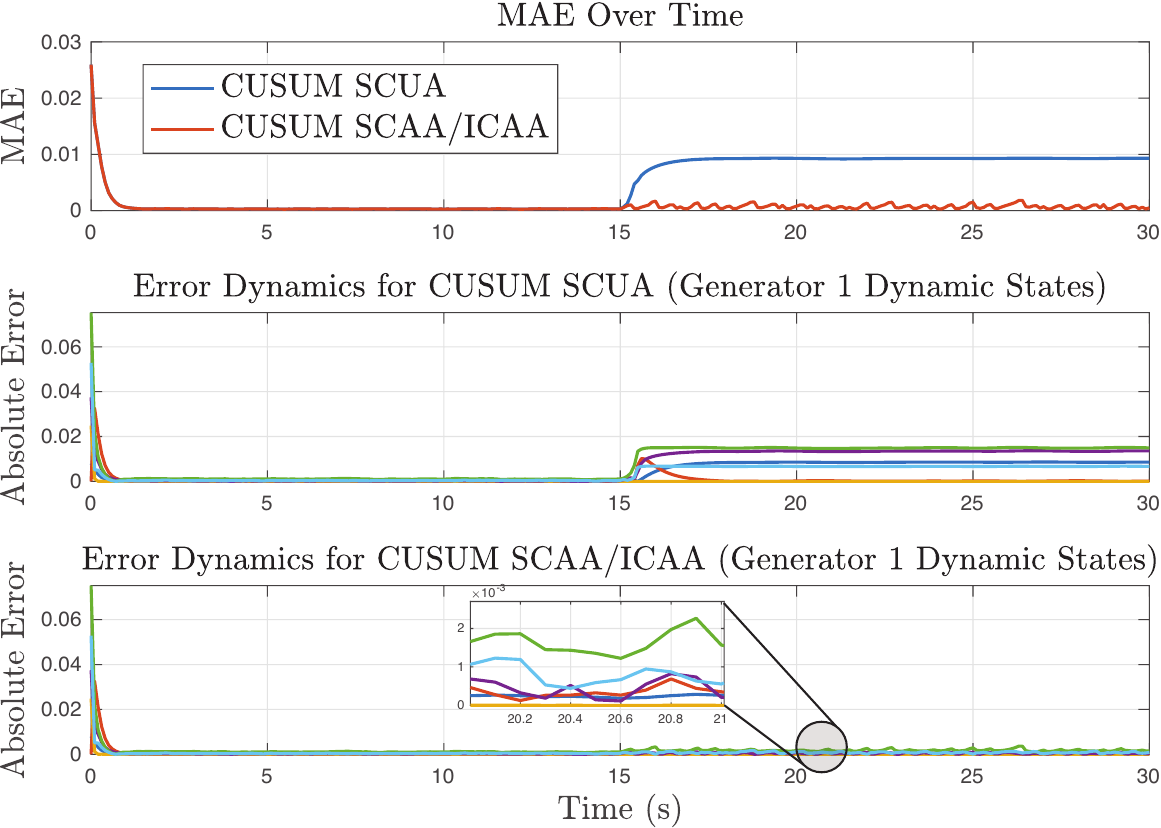}
	\caption{Comparison of the Mean Absolute Error (MAE) and Absolute Error Dynamics for CUSUM SCUA and SCAA/ICAA. }
	\label{fig:MAE}
\end{figure}
\subsection{Algebraic Constraint Violations in SCUAs}
Fig. \ref{fig:violations} illustrates the number of algebraic constraint violations over time for 50 different SCUA attempts. Constraint violations are consistently zero before the attack initiation at t = 15 seconds, confirming the validity of the NDAE model and SE process under normal conditions. Upon attack initiation, there is an immediate spike in violations, this sharp increase demonstrates the SCUA's disregard for NDAE model constraints. After the initial spike, the number of violations generally stabilizes for most attack attempts. This plateau suggests that while SCUAs can maintain their intrusion detection evasion, they consistently violate the underlying physical constraints of the NDAE model.
These results underscore the importance of incorporating NDAE model constraints in attack detection schemes. While SCUAs may evade traditional residual-based detectors, their consistent violation of algebraic constraints provides an additional layer for identifying malicious activities in NDAE-modeled power systems.
\begin{figure}[http]
	\centering
	\includegraphics[width=0.85\columnwidth]{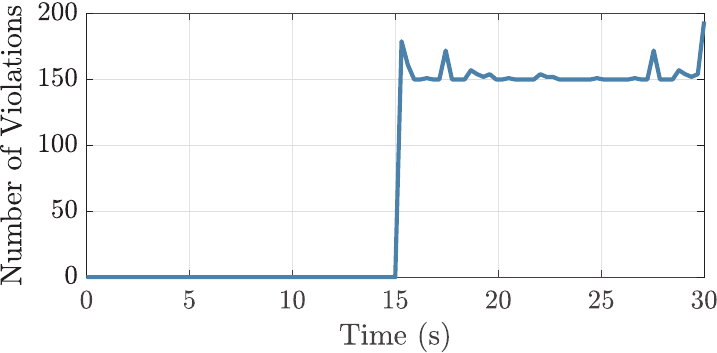}
	\caption{Algebraic constraint violations over time during 50 SCUA attempts using different targetted buses.} 
	\label{fig:violations}
\end{figure}

\subsection{Comparison between SCAA strategies }
The Iterative Constraint-Aware Attack Algorithm (ICAA) (defined in Algorithm \ref{alg:ICAA}) achieves results comparable to the optimization-based SCAA approach \eqref{eq:OP_SFDIA}, and in some cases, it even performs better. The key advantage of ICAA lies in its simplicity. Instead of solving a complex optimization problem, ICAA starts with the maximum possible attack vector and iteratively refines it in small steps to find a vector that bypasses detection, satisfies state estimation, and adheres to all physical constraints of the power system. Algorithm 1 governs this process, focusing only on states in the attack zone defined by Algorithm 2, reducing computational overhead significantly.
\begin{figure}[t]
	\centering
	\includegraphics[width=\columnwidth]{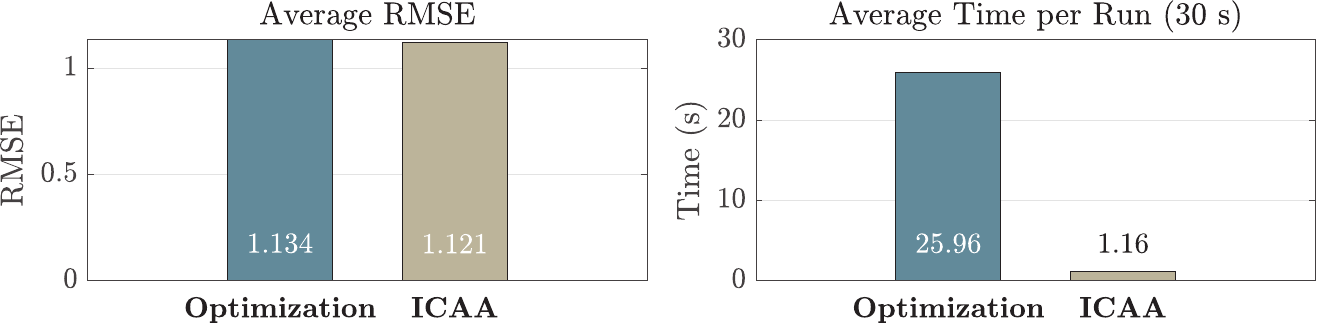}
	\caption{Comparison between SCAA optimization and SCAA-ICAA in 100 differnet runs under chi-squared detection settings} 
	\label{fig:opt_vs_alg}
\end{figure}
As shown in Fig.~\ref{fig:opt_vs_alg}, results from 100 different runs using chi-squared for ID (with varying measurement noise and initial conditions) indicate that the average MAE is almost identical for both approaches. Given these similar outcomes, we opt to use ICAA in the analysis of the attacks in the case studies, where SCAA/ICAA is consistently employed for evaluating attack strategies. This choice highlights the practicality and computational efficiency of ICAA as a robust tool for analyzing stealthy attacks on power systems.

\subsection{Computational Efficiency of SCAA-ICAA}
Tab.~\ref{tab:beta_time_rmse} presents the relationship between the reduction factor ($\beta$), simulation time, and RMSE for SCAA-ICAA attacks over a 30-second period. These results demonstrate the trade-off between attack impact and computational efficiency in the ICAA approach.

\begin{table}[t]
	\centering
\scriptsize	\renewcommand{\arraystretch}{1.2} 
	\caption{Effect of reduction factor ($\beta$) on simulation time and RMSE for SCAA attacks over a 30-second period.}
	\label{tab:beta_time_rmse}
	\begin{tabular}{lcc}
		\toprule
		\textbf{Reduction Factor ($\beta$)} & \textbf{Simulation Time (s)} & \textbf{RMSE} \\
		\midrule
		0.1   & 0.9932  & 0.3571 \\
		0.01  & 2.8884  & 0.3467 \\
		0.001 & 22.6213 & 0.3467 \\
		\bottomrule
	\end{tabular}
\end{table}

A $\beta$ value of (0.1) results in faster computation times (0.9932 seconds) while achieving the highest RMSE (0.3571), indicating the most impactful attack vector. Decreasing $\beta$ to 0.01 increases computation time nearly threefold (2.8884 seconds) while resulting in a slightly lower RMSE (0.3467), suggesting a marginal decrease in attack effectiveness. Further reduction of $\beta$ to 0.001 leads to a significant increase in computation time (22.6213 seconds) without any improvement in RMSE, which remains constant at 0.3467. This indicates that a $\beta$ value of 0.1 provides the optimal trade-off between attack impact and computational efficiency in our NDAE-modeled power system. The lack of significant improvement in attack effectiveness when reducing $\beta$ below 0.1, despite the substantial increase in computation time, suggests that more fine-grained iterations do not yield more potent attacks in this scenario. This behavior likely stems from the inherent constraints of the NDAE model, which limit the extent to which an attacker can manipulate state estimates while maintaining plausibility and evading detection.

\subsection{Sensitivity of SCAAs to Algebraic Constraint Tolerance}
Fig.~\ref{fig:RMSE_vs_zeta} demonstrates the significant impact that the tolerance threshold $\zeta$ for algebraic constraints has on the effectiveness of SCAAs in NDAE power system models. The $\zeta$ parameter represents the maximum allowed deviation from zero for the sum of algebraic constraints $(g(x) = 0)$ in the NDAE model.
\begin{figure}[t]
	\centering
	\includegraphics[width=0.95\columnwidth]{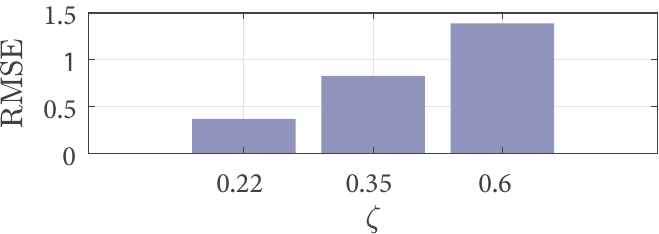}
	\caption{Impact of algebraic constraint tolerance $\zeta$ on Root Mean Square Error (RMSE) of state estimates under Stealthy Constraint-Aware Attacks (SCAAs)} 
	\label{fig:RMSE_vs_zeta}
\end{figure}

The results demonstrate a clear trade-off in NDAE-based power system security. Tighter constraints (lower $\zeta$) significantly limit an attacker's ability to manipulate state estimates without detection, as evidenced by the lower RMSE at $\zeta = 0.22$. However, overly strict constraints may lead to false alarms due to normal system fluctuations or measurement noise. Conversely, looser constraints (higher $\zeta$) reduce false positives but increase vulnerability to sophisticated attacks, as shown by the sharp RMSE increase at $\zeta = 0.6$.

It's important to note that in our simulations, which incorporate loads, renewables, and various noise sources, we observe an average constraint violation of 0.22 under normal conditions. This baseline deviation from zero stems from inherent system uncertainties and estimation limitations. Thus, the chosen $\zeta$ values are calibrated to account for these practical considerations, ensuring meaningful constraint validation in operational settings.

\subsection{Comparison of CUSUM and Chi-squared Detectors}
Fig. \ref{fig:CUSUM_vs_chi} compares the RMSE values for state estimates under SCUA and SCAA-ICAA strategies, considering both CUSUM and chi-squared detectors. Under normal conditions with no attack, both detectors show similar low RMSE values (approximately 0.3), validating the SE accuracy in the absence of attacks. For SCUAs, the CUSUM detector shows an RMSE increase to about 1.3, indicating substantial state estimate deviation, while the chi-squared detector's RMSE rises to approximately 1.56, suggesting even greater vulnerability to SCUAs. In the case of SCAAs, the CUSUM detector shows a moderate RMSE increase to about 0.4, demonstrating the effectiveness of constraint-aware attacks in limiting detectable impacts, while the chi-squared detector's RMSE reaches about 1.3, indicating higher susceptibility to SCAAs compared to CUSUM.

\begin{figure}[t]
	\centering
	\includegraphics[width=0.9\columnwidth]{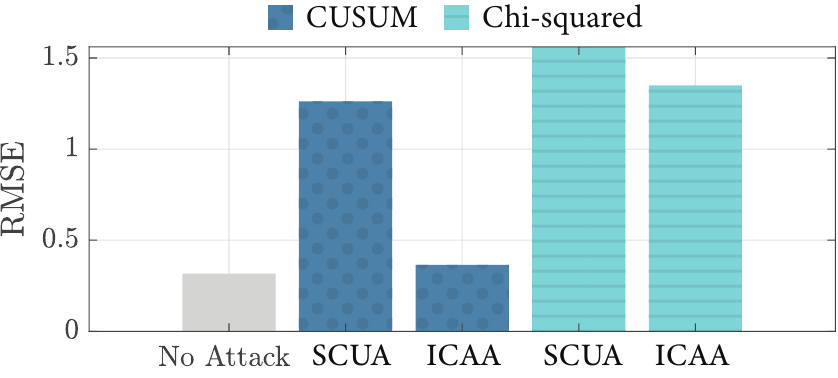}
	\caption{Comparison of CUSUM and chi-squared detection under SCUAs and SCAAs} 
	\label{fig:CUSUM_vs_chi}
\end{figure}
These results highlight several key points. The CUSUM detector generally outperforms the chi-squared detector in limiting the impact of both SCUA and SCAA strategies. This suggests that the CUSUM's ability to track cumulative changes over time provides an advantage in detecting subtle, persistent attacks in NDAE systems. SCAAs consistently produce lower RMSE values compared to SCUAs for both detector types. This demonstrates how incorporating NDAE model constraints significantly restricts the attacker's ability to manipulate state estimates while maintaining stealthiness. The chi-squared detector's higher RMSE values, especially for SCAAs, indicate its limitations in capturing the complex dynamics of NDAE systems. These findings emphasize the need for advanced, NDAE-specific detection methods that leverage both dynamic residual analysis and constraint validation to enhance power system cybersecurity.

\subsection{Aggregated vs. Individual CUSUM Detection}

Fig. \ref{fig:Aggreg_vs_Indiv_CUSUM} presents a comparison of RMSE values for state estimation under SCAAs using \textit{aggregated} (single scalar distance measure) and \textit{individual} (vectorized distance measure) CUSUM detection, across different numbers of compromised measurements (1, 25, and 50). For a single compromised measurement, both detection methods produce similar RMSE values. However, as the number of compromised measurements increases, notable differences emerge.

\begin{figure}[t]
	\centering
	\includegraphics[width=0.95\columnwidth]{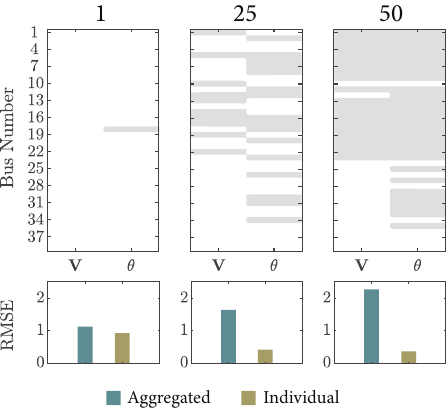}
	\caption{Comparison of RMSE Under SCAA for Aggregated vs. Individual CUSUM Detection} 
	\label{fig:Aggreg_vs_Indiv_CUSUM}
\end{figure}

When using the aggregated CUSUM approach, RMSE progressively rises as more measurements are compromised, reaching a value of approximately 2 when 50 measurements are involved. This suggests that by aggregating residuals, attackers can spread the impact across multiple measurements, keeping the overall detection statistic below the threshold. In contrast, the individual CUSUM approach shows a decrease in RMSE as the number of compromised measurements increases, with RMSE dropping to around 0.5 for 50 measurements. This indicates that monitoring each measurement individually makes it harder for attackers to simultaneously meet all individual thresholds.
This represents a trade-off in CUSUM detector design for NDAE systems. Aggregated approaches, though computationally efficient, may be more susceptible to distributed attacks. On the other hand, individual monitoring offers stronger resistance against large-scale attacks but comes with increased computational demands and a higher likelihood of false alarms.


\subsection{Addressing the Research Questions}
The case studies and analyses presented in this section address the three research questions posed in the introduction.
\begin{itemize}[leftmargin=*]
	\item \textit{A1. Impact of NDAE models on SA formulation, detection, and impact:} NDAE models significantly constrain the feasible attack space due to the coupling between dynamic and algebraic states. SCUAs, while evading detection, consistently violate algebraic constraints, making them detectable through constraint validation. SCAAs, respecting these constraints, have a more limited impact on state estimates. This demonstrates how NDAE models provide inherent advantages over simplified representations by enforcing physical consistency between dynamic and steady-state behaviors.
	\item \textit{A2. Effectiveness of intrusion detectors for NDAE-modeled systems:} CUSUM detectors outperform chi-squared detectors in mitigating both SCUA and SCAA impacts, likely due to their ability to track cumulative changes in the coupled dynamic-algebraic state space. However, both detectors show limitations in fully capturing NDAE dynamics, particularly when attacks target multiple measurements simultaneously. This highlights the need for NDAE-specific detection methods that leverage both temporal residual analysis and instantaneous constraint validation.
	\item  \textit{A3. Performance of NDAE-specific state monitoring algorithms:} Joint state estimation, coupled with dynamic intrusion detection, demonstrates improved resilience against tailored SAs by simultaneously constraining both dynamic and algebraic state estimates. The effectiveness of these methods is highly sensitive to parameter tuning, particularly the algebraic constraint tolerance. Tighter constraints significantly limit an attacker's ability to manipulate state estimates without detection, but may increase false positives due to normal system fluctuations or measurement noise.
\end{itemize}	

	\section{Paper Limitations and Future Work} \label{sec:limit}
	This study presents significant advancements in the understanding of stealthy attacks on NDAE-modeled power systems, yet it has several limitations that warrant further investigation. First, the proposed attack strategies are developed under the assumption that attackers possess complete knowledge of the system topology, model parameters, and intrusion detection mechanisms. While such assumptions are standard in existing literature and allow for a thorough analysis of worst-case scenarios, real-world attackers may operate with partial or incomplete information. Future research should explore how incomplete knowledge affects the design and efficacy of stealthy attacks.\\
	Second, the case studies in this paper are limited to the IEEE 39-bus system using a fourth-order generator model. While this benchmark system effectively demonstrates the feasibility and impact of the proposed methods, it represents a simplified view of real-world power systems. In future work, the proposed methodologies could be extended to larger systems, such as the IEEE 118-bus or 300-bus systems, and incorporate higher-order generator models that capture additional dynamic features, including turbine and governor dynamics. Additionally, future studies could include advanced renewable energy models to reflect the increasing penetration of variable generation sources such as wind and solar power.\\
	Despite these limitations, it is important to note that the scalability of the proposed methods has been rigorously analyzed. The Iterative Constraint-Aware Attack Algorithm (ICAA) is designed to maintain computational efficiency, with complexity scaling linearly with the size of the attack zone, making it applicable to larger and more complex power systems. This ensures that the framework remains practical for real-world applications, even as the size and complexity of the system grow.\\
	Finally, this work highlights the resilience benefits provided by the coupling of dynamic and algebraic states in NDAE models. It also underscores the need for advanced detection methods tailored to the unique dynamics of NDAE systems. Future research should focus on integrating the proposed methods with emerging intrusion detection technologies, assessing their performance under diverse operational scenarios, and further enhancing the robustness of NDAE-based cybersecurity solutions for smart grids.

\section{Conclusions}\label{sec:conclusion}
This study presents an analysis of stealthy false data injection attacks on nonlinear differential algebraic equation (NDAE) models of power networks. Case studies demonstrate that NDAE models, by capturing the coupling between dynamic and algebraic states, inherently provide greater resistance to tailored cyberattacks compared to simplified power system representations.

The use of NDAE models and NDAE-specific state monitoring techniques allows for a more holistic view of stealthy attack impacts by revealing how measurement manipulations propagate through both algebraic steady-state power flow variables and dynamic generator states. This comprehensive approach offers insights that may be overlooked in studies focusing solely on steady-state AC power flow or isolated dynamic state estimation processes.

Our results highlight the importance of incorporating NDAE model constraints in attack detection schemes. The consistent violation of algebraic constraints by certain attack strategies provides an additional layer for identifying malicious activities, even when traditional residual-based detectors fail. The comparison between CUSUM and chi-squared detectors underscores the superiority of dynamic detection methods in NDAE-modeled systems, though both show limitations in fully capturing the system's complex dynamics.

The proposed attack strategies focus on specific vulnerabilities within particular intrusion detectors, which may reduce their generalizability. However, our goal was not to develop universally applicable strategies but to demonstrate how NDAE models inherently offer stronger resilience compared to simplified system representations. Additionally, the assumption of comprehensive system knowledge for attackers may not hold in real-world scenarios. Future work should consider the impact of partial information on the effectiveness of such attacks. While NDAE models provide enhanced accuracy in capturing system dynamics, they come with increased computational complexity. Research into more efficient numerical methods for real-time NDAE-based security assessment would be an important next step.

In conclusion, this work demonstrates the enhanced cybersecurity resilience offered by NDAE models in power systems. It underscores the importance of using comprehensive system representations in security analysis and lays the groundwork for developing more robust, NDAE-specific detection and mitigation strategies for future smart grids.

\bibliographystyle{IEEEtran}
\bibliography{Paper01}

\appendix
\section{NDAE Model Full Description} \label{appendixA}
We consider the fourth-order model of a synchronous generator. The dynamics of a synchronous generator $i \in \mathcal{G}$ can be represented through: \textit{(i)} the following set of ODEs \cite{Kundur2017} that model generator physics, given as follows:
\begin{subequations}\label{eq:1}
	\begin{align}
		&\dot{\delta}_i = \omega_i - \omega_0  \label{eq:1a} \\
		& M_i\dot{\omega}_i = T_{Mi} - P_{Gi} - D_i(\omega_i - \omega_0) \label{eq:1b} \\
		& T'_{d0i}\dot{E'_{qi}} = -\frac{x_{di}}{x'_{di}}E'_{qi} + \frac{x_{di}-x'_{di}}{x'_{di}}v_i\cos(\delta_i - \theta_i) + E_{fdi} \label{eq:1c} \\
		& T'_{q0i}\dot{E'_{di}} = -E'_{di} + \frac{x_{qi}-x'_{qi}}{x_{qi}}v_i\sin(\delta_i - \theta_i),
	\end{align}
\end{subequations}
and \textit{(ii)} the algebraic constraints:
{
	\small
	\begin{subequations}\label{eq: alg}
		\begin{align}
			&P_{Gi} = \frac{1}{x'_{di}} E'_{qi}v_i \sin(\delta_i - \theta_i) - \frac{x_{qi} - x'_{di}}{2x'_{di}x_{qi}}v_i^2 \sin(2(\delta_i - \theta_i)) \label{eq:2a} \\
			&Q_{Gi} = \frac{1}{x'_{di}} E'_{qi}v_i \cos(\delta_i - \theta_i) - \frac{x'_{di} + x_{qi}}{2x'_{di}x_{qi}}v_i^2  \nonumber \\
			&\phantom{=} - \frac{x_{qi} - x'_{di}}{2x'_{di}x_{qi}}v_i^2 \cos(2(\delta_i - \theta_i)) \label{eq:2b} \\
			&P_{Gi} + P_{Ri} - P_{Li} = \sum_{j=1}^{N} v_i v_j (G_{ij} \cos \theta_{ij} + B_{ij} \sin \theta_{ij}) \label{eq:2c} \\
			&Q_{Gi} + Q_{Ri} - Q_{Li} = \sum_{j=1}^{N} v_i v_j (G_{ij} \cos \theta_{ij} - B_{ij} \sin \theta_{ij}) \label{eq:2d},
		\end{align}
	\end{subequations}
}
that consist of the active and reactive power flow equations, shown in (\ref{eq:2a}) and (\ref{eq:2b}), and the power balance equations among generators, loads and renewables, shown in (\ref{eq:2c}) and (\ref{eq:2d}) \cite{Kundur2017}. The system is characterized by dynamic states $\boldsymbol{x}_d\in \mathbb{R}^{n_d}$, including rotor angles $\boldsymbol{\delta}$, angular velocities $\boldsymbol{\omega}$, transient voltages on the q-axis $\boldsymbol{E'_{qi}}$, and transient voltages on the d-axis $\boldsymbol{E'_{di}}$, where $n_d$ is the overall number of dynamic states (equals $4n_g$, where $n_g$ is the number of generators in the system). The algebraic states $\boldsymbol{x}_a\in \mathbb{R}^{n_a}$ comprise generated active and reactive powers $\boldsymbol{P}_G$ and $\boldsymbol{Q}_G$, voltage magnitudes $\boldsymbol{v}$, and angles $\boldsymbol{\theta}$. Parameter $n_a$ defines the number of algebraic states in the system. Inputs and demands are encapsulated in vectors $\boldsymbol{u}\in \mathbb{R}^{n_u}$, representing mechanical torques and field voltage inputs, and $\boldsymbol{w}\in \mathbb{R}^{n_w}$, denoting active and reactive power demands, with $n_u$ and $n_w$ defining the number of control input and uncertain variables. The NDAE model is succinctly represented as:
\begin{subequations}
	\begin{align}
		\textit{Gen. Dynamics:}\;\;	&{\boldsymbol{\dot{x}}}_d = \boldsymbol{A}_d \boldsymbol{x}_d + \boldsymbol{f}_d(\boldsymbol{x}_d, \boldsymbol{x}_a) + \boldsymbol{B}_d \boldsymbol{u}\\
		\textit{Flow Constraints:}\;\;		&\mathbf{0} = \boldsymbol{A}_a \boldsymbol{x}_a +  \boldsymbol{f}_a(\boldsymbol{x}_d, \boldsymbol{x}_a) + \boldsymbol{B}_a \boldsymbol{w}.
	\end{align}
\end{subequations}
The functions \(\boldsymbol{f}_a: \mathbb{R}^{n_d} \times \mathbb{R}^{n_a} \to \mathbb{R}^{n_{f a}}\), and \(\boldsymbol{f}_d: \mathbb{R}^{n_d} \times \mathbb{R}^{n_a} \to \mathbb{R}^{n_{f d}}\) describe the nonlinearities in the algebraic and dynamic states respectively. The state-space matrices \(\boldsymbol{A}_a \in \mathbb{R}^{n_a \times n_a}\), \(\boldsymbol{B}_a \in \mathbb{R}^{n_a \times n_q}\), \(\boldsymbol{A}_d \in \mathbb{R}^{n_d \times n_d}\), \(\boldsymbol{B}_d \in \mathbb{R}^{n_d \times n_u}\) are all detailed in~\cite{Nugroho2022a}. The model can be more concisely expressed in this standard nonlinear DAE form:
\begin{equation}
	\boldsymbol{E} \dot{\boldsymbol{x}} = \boldsymbol{A} \boldsymbol{x} + \boldsymbol{f}(\boldsymbol{x}) + \boldsymbol{B}_u \boldsymbol{u} +\boldsymbol{B}_w \boldsymbol{w},
\end{equation}
where the complete state vector $\boldsymbol{x}$ combines both dynamic and algebraic states. This form is facilitated by:
\begin{align*}
	\boldsymbol{E} &= \begin{bmatrix}
		\boldsymbol{I} & \boldsymbol{O} \\
		\boldsymbol{O} & \boldsymbol{O}
	\end{bmatrix}, \boldsymbol{A} = \begin{bmatrix}
		\boldsymbol{A}_d & \boldsymbol{O} \\
		\boldsymbol{O} & \boldsymbol{A}_a
	\end{bmatrix}, \
	\boldsymbol{f}(\boldsymbol{x}) = \begin{bmatrix}
		\boldsymbol{f}_d(\boldsymbol{x}) \\
		\boldsymbol{f}_a(\boldsymbol{x})
	\end{bmatrix}, \\
	\boldsymbol{B}_u &= \begin{bmatrix}
		\boldsymbol{B}_d \
		\boldsymbol{O}
	\end{bmatrix}^{\top},  \boldsymbol{B}_w = \begin{bmatrix}
		\boldsymbol{O} \
		\boldsymbol{B}_a
	\end{bmatrix}^{\top},
\end{align*}
where matrix $\m E$ is singular, thereby encoding the algebraic flow constraints.

\section{Attack Zone Algorithm} \label{appendixB}}

	\begin{algorithm}[http] 		

		\caption{Attack Zone Definition}
			\label{algo:zone}
		\begin{algorithmic}[1]
			\Require Target buses $\mathcal{T}$, network admittance matrix $\boldsymbol{Y}$, bus classifications, depth limit $d_{max}$
			\Ensure Attack zone $\mathcal{A}$, required state indices $\mathcal{S}$
			\State Classify buses into zero-injection $\mathcal{Z}$ and non-zero injection $\mathcal{N}$ sets
			\State Initialize $\mathcal{A} \gets \mathcal{T}$
			\State Initialize queue $Q \gets \mathcal{T}$, visited $V \gets \emptyset$, depth $d \gets 0$
			\While{$Q \neq \emptyset$ \textbf{and} $d < d_{max}$}
			\State $Q_{next} \gets \emptyset$
			\For{each bus $i \in Q$}
			\State Find connected buses $\mathcal{C}_i \gets {j : |Y_{ij}| > \epsilon}$
			\State $\mathcal{C}_i \gets \mathcal{C}_i \setminus V$
			\For{each bus $j \in \mathcal{C}i$}
			\If{$j \in \mathcal{Z}$}
			\State $\mathcal{A} \gets \mathcal{A} \cup {j}$
			\State $Q_{next} \gets Q_{next} \cup {j}$
			\ElsIf{$j \in \mathcal{N}$}
			\State $\mathcal{A} \gets \mathcal{A} \cup {j}$
			\EndIf
			\EndFor
			\State $V \gets V \cup {i}$
			\EndFor
			\State $Q \gets Q_{next}$
			\State $d \gets d + 1$
			\EndWhile
			\State Find all connected buses $\mathcal{B} \gets {j : \exists i \in \mathcal{A}, |Y_{ij}| > \epsilon}$
			\State Initialize $\mathcal{S} \gets \emptyset$
			\For{each generator bus $i \in \mathcal{G}$}
			\If{$i \in \mathcal{A} \cup \mathcal{B}$}
			\State Add generator states to $\mathcal{S}$
			\State Add bus voltage and angle states to $\mathcal{S}$
			\EndIf
			\EndFor
			\For{each non-generator bus $i \in \mathcal{A} \cup \mathcal{B}$}
			\State Add bus voltage and angle states to $\mathcal{S}$
			\EndFor
			\State \Return Attack zone $\mathcal{A}$, state indices $\mathcal{S}$
		\end{algorithmic}
	\end{algorithm}
\end{document}